\documentclass[pra,aps,twocolumn,showpacs]{revtex4}
\usepackage{bm}
\usepackage{graphicx}

\begin{document}
\title{Lower bounds on the absorption probability of beam splitters}
\author{S.~Scheel}
\email{s.scheel@imperial.ac.uk}
\affiliation{Quantum Optics and Laser Science, Blackett Laboratory,
Imperial College London, Prince Consort Road, London SW7 2BW, United Kingdom}

\date{\today}

\begin{abstract}
We derive a lower limit to the amount of absorptive loss present in
passive linear optical devices such as a beam splitter. We choose a
particularly simple beam splitter geometry, a single planar slab
surrounded by vacuum, which already reveals the important features of
the theory. It is shown that, using general causality requirements and
statistical arguments, the lower bound depends on the frequency of the
incident light and the transverse resonance frequency of a suitably
chosen single-resonance model only. For symmetric beam splitters and
reasonable assumptions on the resonance frequency $\omega_T$, the
lower absorption bound is $p_{\min}\approx 10^{-6}(\omega/\omega_T)^4$.
\end{abstract}


\maketitle
%%%%%%%%%%%%%%%%%%%%%%%%%%%%%%%%%%%%%%%%%%%%%%%%%%%%%%%%%%%%%%%%%%%%%%
\section{Introduction}

Passive linear optical elements such as beam splitters are
indispensible tools in photonic interferometry. Most recently, they 
became even more important with the emergence of ideas to use beam
splitters as central units in all-optical quantum information
processing \cite{KLM}. The action of a beam splitter can be understood
by assuming that light impinges on a single slab of material with a
refractive index $\eta$. The differences in refractive indices at the
faces of the beam splitter cause light to be partially reflected from
the surface, while other parts are transmitted.

In a mathematical description of it, the beam splitter is mostly
assumed to be lossless, at least in the frequency window associated
with the band width of the incident light. In this case, the action of
a beam splitter on two amplitude operators of photons impinging on it
is given by an element of the unitary group SU(2) which reflects
energy conservation by splitting into transmitted and reflected light
\cite{lossless}.
If two identical photons in well-defined
single-photon Fock states impinge on both sides of a lossless beam
splitter, one would for instance observe the well-known
Hong--Ou--Mandel quantum interference effect \cite{HOM} with perfect
visibility.

Realistically, however, perfectly lossless beam splitter do not exist
and cannot exist. Causality and the resulting Kramers--Kronig
relations for the dielectric permittivity imply that a (real part of
the) refractive index $\eta(\omega)$ different from unity is always
accompanied by an imaginary part $\kappa(\omega)$ responsible for
describing absorption and dissipation \cite{Nussenzveig}. An
additional feature of the so-called superconvergence rule
\cite{Altarelli72} which follows from the Kramers--Kronig relations is
that the real part of the refractive index is unity on average over
all frequencies, i.e. $\int_0^\infty d\omega[\eta(\omega)-1]=0$.

On the other hand, causality itself does not immediately imply that
the absorption coefficient has to be strictly larger than zero for all
non-zero frequencies. In fact, the Kramers--Kronig relations in
principle allow for a situation in which there is only one $\delta$
function peak in the absorption spectrum at some particular frequency
and vanishing absorption elsewhere. Physically, this implies an
infinitely sharp resonance and thus an infinitely long lifetime of the
corresponding transition which is not realistic either. Here
some statistical arguments come to the rescue. For macroscopically
large systems a statistical description in terms of response functions
has to be used which leads to a particular form of the absorption
spectrum. In the linear-response approximation, this prevents the
absorption function to become zero at any point except the origin of
the frequency spectrum.

A causal response of a macroscopic physical quantity (such as the
dielectric polarization) to an external excitation (such as an
electric field) results in relations between the correlation function
for the responding quantities and the imaginary part of the response
function. Such relations are known as fluctuation-dissipation theorems
\cite{CallenWelton}. They have wide applications in all branches of
statistical physics. Of particular interest to us is that
fluctuation-dissipation theorems have helped to establish limits on
physical processes such as frequency stabilization of lasers with
cavities \cite{Numata04}. Generalizations of
these theorems have been used to provide bounds on
the performance of solid-state two-qubit gates \cite{Fisher}. 

A realistic description of a beam splitter that includes absorption
and dispersion has been formulated in \cite{Knoll99} and makes use of
the quantum-optical input-output relations derived in
\cite{Gruner96}. It is shown in \cite{Knoll99} that the action of a 
lossy beam splitter on the amplitude operators of the incoming light
(plus the amplitude operators of the device excitation) corresponds to
an element of the group SU(4). From there the SU(2)-action of a
lossless beam splitter follows as a limiting case in which photons and
device would effectively decouple which, by the above arguments, means
that there is no beam splitter action present at all.

The consequence of this argument is that any beam splitter, no matter
how well it is fabricated, must show some non-vanishing
absorption. In this article we will show how to obtain lower bounds on
the amount of absorption. We will use a simple model of a single
planar slab for which the input-output relations are well-known
\cite{Gruner96} and we will use a single-resonance Drude--Lorentz
model for the dielectric permittivity (Sec.~\ref{sec:model}). This,
together with some lower bounds on the resonance line width, will be
sufficient to derive such lower bounds that only depend on the
frequency of the impinging light (Sec.~\ref{sec:limit}). We summarize
our results in Sec.~\ref{sec:conclusions}.

%%%%%%%%%%%%%%%%%%%%%%%%%%%%%%%%%%%%%%%%%%%%%%%%%%%%%%%%%%%%%%%%%%%%%%
\section{The beam splitter model}
\label{sec:model}

We will use a particularly simple model for a beam splitter, a single
planar slab of thickness $l$ made of a material with a complex
refractive index $n(\omega)=\eta(\omega)+i\kappa(\omega)$. In this
approximation, we consider linearly polarized light travelling in a
particular direction in space. Following \cite{Gruner96}, the
transmission and reflection coefficients $T(\omega)$ and $R(\omega)$,
respectively, can be written in the form 
\begin{eqnarray}
\label{eq:trans}
T(\omega) &=&
\frac{4n e^{i(n-1)\omega l/c}}{(1+n)^2-(1-n)^2 e^{2in\omega l/c}}
\,,\\ \label{eq:ref}
R(\omega) &=& \frac{n-1}{n+1} e^{-i\omega l/c}
\left[ 1-T(\omega) e^{i(n-1)\omega l/c} \right] \,,
\end{eqnarray}
[$n\equiv n(\omega)$]. Because of absorption, they fulfil
$|T(\omega)|^2+|R(\omega)|^2\le 1$.

In the next step, we need to model the complex refractive index
$n(\omega)$. In the linear approximation in which the macroscopic
polarization responds linearly to an external electric field, we need to
define a complex susceptibility function $\chi(\omega)$. In this
approximation the quasi-excitations associated with the linear
response are harmonic oscillator-like, hence the susceptibility can
only be a (possibly infinite) sum of Drude--Lorentz functions, i.e.
\begin{equation}
\label{eq:chi}
\chi(\omega) = \sum\limits_i
\frac{\omega_{P,i}^2}{\omega_{T,i}^2-\omega^2-i\gamma_i\omega} \,,
\end{equation}
where $\omega_{T,i}$ is the resonance frequencies of the $i$th
resonance with a decay constant or line width $\gamma_i$, and
strengths $\omega_{p,i}$ that are related to the static dielectric
permittivity. Note that Drude--Lorentz susceptibilities are the only
allowed type of response functions. From Eq.~(\ref{eq:chi}) it is seen
that the susceptibility has a strictly positive imaginary part for all
positive frequencies. That means that even at frequencies far away
from all resonances, a beam splitter will absorb light and we have the
strict inequality $|T(\omega)|^2+|R(\omega)|^2< 1$ except for the
point $\omega=0$. In particular, for frequencies below all resonances,
we can write for the real and imaginary parts of the refractive index
to their respective lowest orders in $\omega$
\begin{eqnarray}
\eta(\omega) &\approx& \left( 1+\sum\limits_i
\frac{\omega_{P,i}^2}{\omega_{T,i}^2} \right)^{1/2} \,, \\
\kappa(\omega) &\approx& \frac{\omega}{2\eta(\omega)}
\sum\limits_i \frac{1}{\omega_{T,i}}
\left( \frac{\gamma_i}{\omega_{T,i}}\right)
\left( \frac{\omega_{P,i}^2}{\omega_{T,i}^2} \right) \,.
\end{eqnarray}
This limit is of special importance for our consideration since it is
the regime in which absorption is lowest while $n(\omega)$ retains a
strong diffractive part.

\section{Minimal absorption probability}
\label{sec:limit}

Let us assume that we would like to build a beam splitter with a
certain splitting ratio $x=|T|^2/|R|^2$ out of such a single
slab. Then there are several parameters which we can tune: the static
permittivity, the frequency of the incoming light, the thickness of
the slab, and the line widths of the resonances. 
If we restrict ourselves to a single-resonance model with a transition
frequency $\omega_T$, line width $\gamma$, and strength $\omega_P$, we
are left with four parameters: the static refractive index
$\eta=1+\omega_P^2/\omega_T^2$, the scaled line width
$\tilde{\gamma}=\gamma/\omega_T$, the scaled frequency
$\tilde{\omega}=\omega/\omega_T$, and the 
scaled thickness $d=\omega_T l/c$. The product $\eta d$ is
proportional to the optical path length and thus determined by the
chosen beam splitting ratio $x$. Then, given the frequency
$\tilde{\omega}$ and the line width $\tilde{\gamma}$, we minimize the
absorption probability $p=1-|T|^2-|R|^2$ over the beam splitter
thickness $d$. For low frequencies and narrow line widths we can expand
Eqs.~(\ref{eq:trans}) and (\ref{eq:ref}) with respect to those small
parameters. The result can be cast in the form
\begin{equation}
\label{eq:pmin}
p_{\min} = \alpha(x) \tilde{\gamma} \tilde{\omega} \,,
\end{equation}
where $\alpha(x)$ is a numerical coefficient that depends on the beam
splitting ratio $x$.
\begin{figure}[ht] 
\includegraphics[width=8cm]{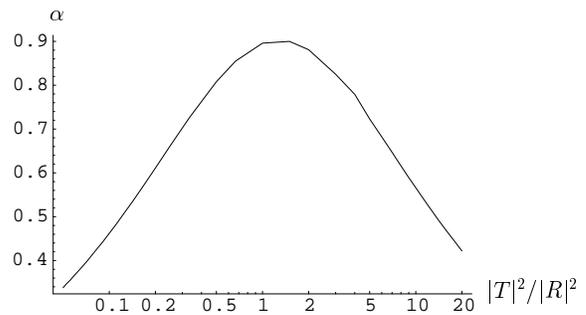}
\caption{\label{fig:bild} Coefficient \protect$\alpha$ in
Eq.~(\ref{eq:pmin}) for different beam splitting ratios
$x=|T|^2/|R|^2$. The maximum is obtained around a symmetric splitting
ratio.}
\end{figure}
Figure~\ref{fig:bild} shows the result of a numerical calculation of
the coefficient $\alpha(x)$ for different $x$. The maximum of the
curve is obtained around the symmetric splitting ratio $x=1$. At this
point, $\alpha(1)\approx 0.9$. In the extreme cases when either $x\to
0$ (perfect reflection) or $x\to\infty$ (perfect transmission),
$\alpha(x)$ tends to zero. However, in the case $x\to 0$ that does not
say that the absorption probability vanishes.

The minimal absorption probability (\ref{eq:pmin}) still depends on
two parameters, the resonance line width $\tilde{\gamma}$ and the
frequency of the incoming light, $\tilde{\omega}$. In order to find
the absolute minimum of Eq.~(\ref{eq:pmin}) we need to give a lower
bound on the scaled line width $\tilde{\gamma}$.
For a non-vanishing susceptibility we need non-zero dipole moment
which implies the existence of some (electronic) excited states that
can be populated by optical excitation. This is because the dipole
moment is proportional to the off-diagonal matrix element of the
position operator. These electronic states are coupled to the
electromagnetic vacuum and are thus subject to spontaneous
decay. Disregarding fabrication errors and impurities, spontaneous
decay is the limiting factor when determining the line width. This
fundamental process is also responsible for giving lower bounds on
factorization times in quantum computers \cite{Plenio}.

Because each individual dipole (atom) is surrounded by many other dipoles
of the same kind, it will feel not only the vacuum fluctuations but
enhanced fluctuations due to the presence of the surroundings. These
effects are known as local-field corrections, and several models exist
to calculate them. In particular, the so-called real-cavity model has
attracted much attention because it consistently places the radiating
dipoles in vacuum \cite{GlauberLewenstein,Scheel99b} rather than in
front of a smeared dielectric background which is known to cause
spurious divergences \cite{Barnett96,Scheel99a,Fleischhauer99}.

Because we are interested in the low-frequency limit in which
absorption is very small, we can ignore its influence on spontaneous
decay and approximate the effect of the surrounding atoms by the
(classical) real-cavity correction factor
\cite{GlauberLewenstein,Scheel99b},
\begin{equation}
\label{eq:gamma}
\Gamma = \eta(\omega) \Gamma_0 \left(
\frac{3\eta^2(\omega)}{2\eta^2(\omega)+1} \right)^2 \,,
\end{equation}
where $\Gamma_0$ is the free-space decay rate which is defined
as
\begin{equation}
\label{eq:gamma0}
\Gamma_0 = \frac{\omega^3d^2}{3\pi\hbar\varepsilon_0 c^3} \,,
\end{equation}
where $d$ is the (matrix element of the) dipole moment of the
transition. The dipole moment is related to the static refractive
index in the following way. Consider a two-level atom with a ground
state $|g\rangle$ and an excited state $|e\rangle$ (there is
no need to consider more complicated situations because the
single-resonance model corresponds directly to the response of an
ensemble of two-level atoms). The atomic polarizability in the
zero-frequency limit can be written as \cite{Melrose}
\begin{equation}
\label{eq:polarizability}
\alpha_{ij}(0) = \frac{n_g-n_e}{\hbar\omega_T nV} \left[ d_id_j^\ast
+d_i^\ast d_j \right] \approx
\frac{d_id_j^\ast +d_i^\ast d_j}{\hbar\omega_T}
%\left[ d_id_j^\ast +d_i^\ast d_j \right]
\end{equation}
where $n_{g,e}$ are the population numbers of ground-state and
excited-state atoms, respectively, and $n$ the total number density of
atoms in the volume $V$. The matrix elements of the dipole operator
are denoted as $d_i=\langle e|\hat{d}_i|g\rangle$. The approximation
made on the rhs of Eq.~(\ref{eq:polarizability}) is justified because
at room temperature and optical frequencies the thermal occupation of
the excited state is negligble, and in the linear-response approximation
there are no population inversions. Therefore, the zero-frequency
susceptibility can be written as 
\begin{equation}
\chi(0) \approx \frac{2d^2 n}{3\hbar\omega_T\varepsilon_0} \,.
\end{equation}
In that way, we can relate the dipole moment to the static refractive
index $\eta$ as
\begin{equation}
\label{eq:dipolemoment}
d^2 = \frac{3\hbar\omega_T\varepsilon_0(\eta^2-1)}{2n} \,.
\end{equation}
Inserting Eq.~(\ref{eq:dipolemoment}) into (\ref{eq:gamma0}) and
subsequently into Eq.~(\ref{eq:gamma}), we find that
\begin{equation}
\label{eq:gamma2}
\Gamma = \frac{4\pi^2}{nV_T} \tilde{\omega}^3 \eta (\eta^2-1) \left(
\frac{3\eta^2}{2\eta^2+1} \right)^2 \,,
\end{equation}
where $V_T$ is the volume of a cube with side length equal to the
transition wavelength $\lambda_T$. The product $nV_T$ is therefore the
number of atoms within this cube.

\begin{figure}[ht]
\includegraphics[width=8cm]{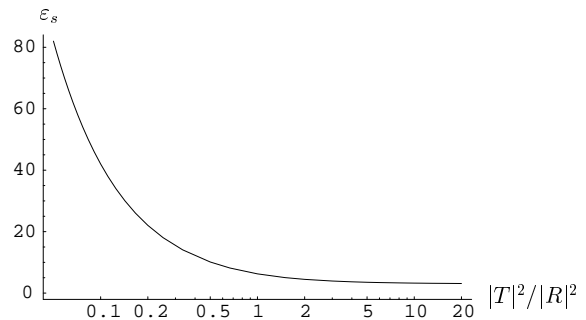}
\caption{\label{fig:epsilon} Static permittivity $\varepsilon_s$
leading to the minimal value $p_{\min}$ [Eq.~(\ref{eq:pmin})] for
different beam splitter ratios $x=|T|^2/|R|^2$.}
\end{figure}

Finally, we associate Eq.~(\ref{eq:gamma2}) with the lower bound on
the line width $\gamma$ of the response function. Hence, we obtain the
following formula for the minimal absorption probability as
\begin{equation}
\label{eq:final}
p_{\min} = \frac{4\pi^2\alpha(x)}{nV_T} \tilde{\omega}^4 \eta (\eta^2-1)
\left( \frac{3\eta^2}{2\eta^2+1} \right)^2 \,.
\end{equation}
Note that the static refractive index $\eta$ appearing in
Eq.~(\ref{eq:final}) has to obtained from the minimization procedure
over the beam splitter thickness $d$ that led to the numerical
coefficient $\alpha(x)$ in Fig.~\ref{fig:bild}. The dependence of the
static permittivity $\varepsilon_s=1+\chi(0)=\eta^2$ on the beam
splitter ratio $x$ is shown in Fig.~\ref{fig:epsilon}. 

As an example, let us consider the symmetric beam splitter with
$x=1$. The numerical minimization leads to $\alpha\approx 0.9$ at
$\varepsilon_s\approx 6.2$. Let us further assume that the resonance
frequency $\omega_T$ lies somewhere in the UV such that we can take
$nV_T\approx 10^9$. With these numbers, the minimal absorption
probability reads
\begin{equation}
p_{\min} \approx 10^{-6} \tilde{\omega}^4 \,.
\end{equation}
If additionally the incident light has a frequency that lies in the IR
region, i.e. $\tilde{\omega}\approx 0.1$, then we obtain
$p_{\min}\approx 10^{-10}$. Currently, the best beam splitters
that have been fabricated show an absorption probability of around
2~ppm \cite{Schnabel}. This means that on the one hand that there is
still room for improvement by eliminating technical imperfections such as
impurities or diffuse scattering from interfaces. On the other hand,
the lower bound is large enough to become an obstacle in future
generations of beam splitter fabrication. 

Equation~(\ref{eq:final}) shows that the minimal absorption
probability strongly depends on the static permittivity (or,
equivalently, the static refractive index). For large values of
$\eta$, $p_{\min}$ increases as $p_{\min}\propto\varepsilon_s^{3/2}$
or $p_{\min}\propto\eta^3$, respectively. From Fig.~\ref{fig:epsilon}
it follows that this is the case for small beam splitting ratios,
i.e. when the slab is highly reflecting. For example, choosing
$x=0.05$ and all other parameters as before, we obtain
$p_{\min}\approx 2\cdot 10^{-5}\tilde{\omega}^4$, an increase by a
factor of 20.

%%%%%%%%%%%%%%%%%%%%%%%%%%%%%%%%%%%%%%%%%%%%%%%%%%%%%%%%%%%%%%%%%%%%%%
\section{Conclusions}
\label{sec:conclusions}

We have shown that one can obtain lower bounds on the absorption
probability of a beam splitter by causality arguments. This shows the
intimate relation between manipulability of light --- described by the
real part of the linear response function --- and decoherence or
absorption which is described by the imaginary part of the response
function. Although it is intuitively clear that such relations must
exist, we have presented in a particularly simple example how they
emerge in a realistic physical situation.

From the Kramers--Kronig relations one infers that a beam splitter
material which necessarily has $\eta(\omega)\ne 1$ shows absorption
somewhere in the frequency spectrum. However, it is only the assumption
of linear response and the resulting validity of Drude--Lorentz models
to describe the dielectric susceptibility that leads to the absorption
coefficient being strictly positive for positive frequencies. In the
single-resonance Drude--Lorentz model employed here the width of the
resonance is limited by the spontaneous decay rate between excited and
ground state. This is certainly a lower limit when all line-broadening
effects are disregarded.

The final result, Eq.~(\ref{eq:final}), is seen to depend only on the
transverse resonance frequency $\omega_T$ and the frequency of the
incident light $\omega$. The other parameters (beam splitter thickness
$l$, static susceptibility $\varepsilon_s$, and line width $\gamma$)
are fixed by the beam splitter condition ($l$), the minimization
($\varepsilon_s$), and the spontaneous decay rate ($\gamma$),
respectively. For reasonable assumptions on the transverse resonance
frequency, we obtain that the absorption limit for a symmetric beam
splitter is approximately
$p_{\min}\approx 10^{-6}\tilde{\omega}^4$. This result will set a
possible limit on experimental performances where ultra-low absorption
is necessary, such as in gravitational-wave interferometry. Currently,
absorption probabilities of 2~ppm have already been achieved, and with
progress being made in eliminating material impurities, surface
roughnesses, and other experimental imperfections it is reasonable to
assume that the theoretical limits of the kind that have been
considered here will impose ultimate performance limitations that have
to be taken seriously.

\acknowledgments
We would like to thank M.B.~Plenio for discussions.
This work was supported by the UK Engineering and Physical Sciences
Research Council (EPSRC).
%%%%%%%%%%%%%%%%%%%%%%%%%%%%%%%%%%%%%%%%%%%%%%%%%%%%%%%%%%%%%%%%%%%%%%

\end{document}